\pdfoutput=1
\documentclass[11pt]{article}

\usepackage{ACL2023}

\usepackage{times}
\usepackage{latexsym}

\usepackage[T1]{fontenc}
\usepackage[utf8]{inputenc}

\usepackage{microtype}

\usepackage{inconsolata}
\usepackage {amsmath}
\usepackage{graphicx}
\usepackage{multicol}
\usepackage{multirow}
\usepackage{subfigure}
\usepackage{stfloats}
\usepackage{color}
\usepackage{subfigure}
 
\definecolor{midnightgreen}{rgb}{0.0, 0.29, 0.33}

\title{Structure-Aware Language Model Pretraining Improves Dense Retrieval on Structured Data}

\author{Xinze Li$^{1}$, Zhenghao Liu$^{1}$\thanks{ \ \ indicates corresponding author.}, Chenyan Xiong$^{2}$, Shi Yu$^{3}$, Yu Gu$^{1}$, Zhiyuan Liu$^{3}$ and Ge Yu$^{1}$ \\ 
$^1$Department of Computer Science and Technology, Northeastern University, China \\
$^2$Microsoft Research, United States \\
$^3$Department of Computer Science and Technology, Institute for AI, Tsinghua University, China \\
Beijing National Research Center for Information Science and Technology, China \\
}

\begin{document}
\maketitle
\begin{abstract}
This paper presents \textbf{S}tructure \textbf{A}ware De\textbf{N}se Re\textbf{T}riev\textbf{A}l (SANTA) model, which encodes user queries and structured data in one universal embedding space for retrieving structured data. SANTA proposes two pretraining methods to make language models structure-aware and learn effective representations for structured data: 1) Structured Data Alignment, which utilizes the natural alignment relations between structured data and unstructured data for structure-aware pretraining. It contrastively trains language models to represent multi-modal text data and teaches models to distinguish matched structured data for unstructured texts. 2) Masked Entity Prediction, which designs an entity-oriented mask strategy and asks language models to fill in the masked entities. Our experiments show that SANTA achieves state-of-the-art on code search and product search and conducts convincing results in the zero-shot setting. SANTA learns tailored representations for multi-modal text data by aligning structured and unstructured data pairs and capturing structural semantics by masking and predicting entities in the structured data. All codes are available at \url{https://github.com/OpenMatch/OpenMatch}.
%Dense retrieval performs well on free-form natural language text retrieval by matching in the learned embedding space. But only free-form text can't satisfy our needs because structured text and free-form text often coexist in our life such as code search and shopping search tasks which require reasoning over both free-form questions and structured text. Besides, existing pre-trained language model cannot learn the representation of structured data well, which makes the dense retrieval for structured data unsatisfactory. In this paper, we present xxx(name), a pre-trained LM that jointly learns representations for NL and structured data to improve the ability of dense retrieval for structured data. For enhancing semantic representation learning of structured data, we use cross-modal text alignment to train the model on the structure data set. Besides, in order to better capture structure information, we mask the some tokens and use model complete them. Finally... Our model beats previous PLM on multiple structure data retrieval. 
\end{abstract}
\section{Introduction}
Dense retrieval has shown strong effectiveness in lots of NLP applications, such as open domain question answering~\cite{chen2017reading}, conversational search~\cite{qu2020orqa,Yu2021FewShotCD}, and fact verification~\cite{thorne2018fact}. It employs pretrained language models (PLMs) to encode unstructured data as high-dimensional embeddings, conduct text matching in an embedding space and return candidates to satisfy user needs~\cite{xiong2020dense,karpukhin2020dense}. 
\begin{figure}[t]
    \centering
    \includegraphics[width=\linewidth]{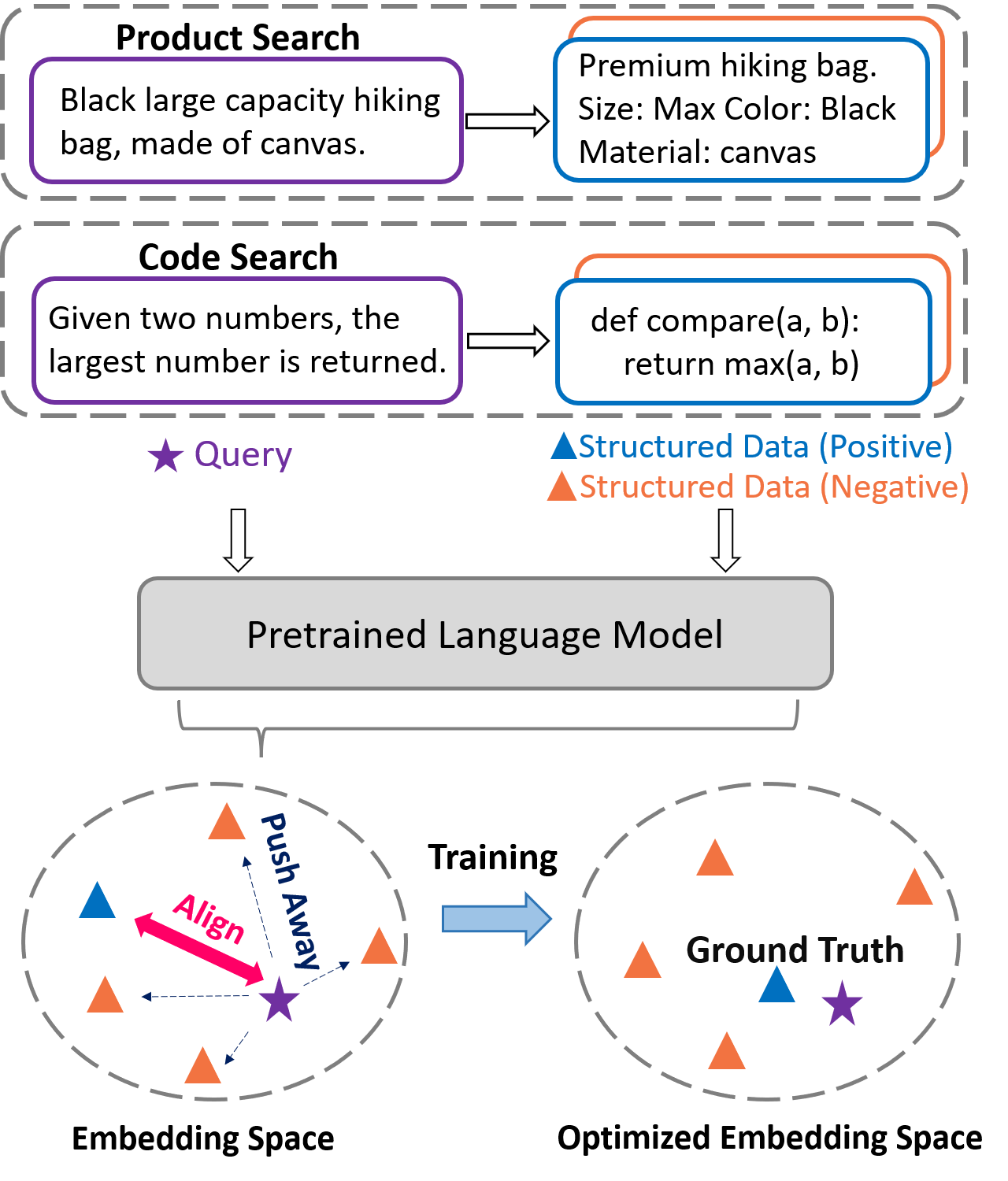}
    \caption{Dense Retrieval Pipeline on Structured Data.}
    \label{fig:case:task}
\end{figure}

Besides unstructured data, structured data, such as codes, HTML documents and product descriptions, is ubiquitous in articles, books, and Web pages, and plays the same important roles in understanding text data. Learning the semantics behind text structures to represent structured data is crucial to building a more self-contained retrieval system. The structured data modeling stimulates researchers to build several benchmarks to evaluate model performance, such as code search and product search~\cite{DBLP:journals/corr/abs-1909-09436,DBLP:journals/corr/abs-2206-06588}. The structured data retrieval tasks require models to retrieve structured data according to user queries. Dense retrieval~\cite{karpukhin2020dense,li2022coderetriever} shows a promising way to build a retrieval system on structured data by encoding user queries and structured data in an embedding space and conducting text matching using the embedding similarity. Nevertheless, without structure-aware pretraining, most PLMs lack the necessary knowledge to understand structured data and conduct effective representations for retrieval~\cite{feng2020codebert,HuYXLLY22,gururangan2020don}.

Lots of structure-aware pretraining methods are proposed to continuously train PLMs to be structure-aware and better represent structured data~\cite{wang2021codet5,feng2020codebert}. They design task-specific masking strategies and pretrain PLMs with mask language modeling. Nevertheless, only using mask language modeling may not sufficiently train PLMs to conduct effective representations for structured data~\cite{li2020sentence,fang2020cert}. Some natural alignment signals between structured and unstructured data, such as code-description documentation and product description-bullet points, provide an opportunity to pretrain the structured data representations. Using these alignment signals, PLMs can be contrastively trained~\cite{wu2020clear,karpukhin2020dense} to match the representations of aligned structured and unstructured data and understand the semantics of structured data with the help of natural language.

In this paper, we propose \textbf{S}tructure \textbf{A}ware De\textbf{N}se Re\textbf{T}riev\textbf{A}l (SANTA), a dense retrieval method on structured data. As shown in Figure~\ref{fig:case:task}, SANTA encodes queries and structured data in an embedding space for retrieval. SANTA designs two pretraining tasks to continuously train PLMs and make PLMs sensitive to structured data. The Structured Data Alignment task contrastively trains PLMs to align matched structured-unstructured data pairs in the embedding space, which helps to represent structured data by bridging the modality gap between structured and unstructured data. The Masked Entity Prediction task masks entities and trains PLMs to fill in the masked parts, which helps to capture semantics from structured data.

Our experiments show that SANTA achieves state-of-the-art in retrieving structured data, such as codes and products. By aligning structured and unstructured data, SANTA maps both structured and unstructured data in one universal embedding space and learns more tailored embeddings for multi-modal text data matching. The masked entity prediction task further guides SANTA to capture more crucial information for retrieval and better distinguish structured and unstructured data. Depending on these pretraining methods, SANTA can even achieve comparable retrieval results with existing code retrieval models without finetuning, showing that our structure-aware pretraining can benefit structured data understanding, multi-modal text data representation modeling and text data matching between user queries and structured data.

\section{Related Work }
Dense retrieval~\cite{Yu2021FewShotCD,karpukhin2020dense,xiong2020dense,li2021more} encodes queries and documents using pretrained language model (PLM)~\cite{devlin2019bert,liu2019roberta,raffel2020exploring} and maps them in an embedding space for retrieval. %Leaning more effective representations for queries and documents with PLMs is crucial for dense retrieval~\cite{cocondenser,luan2020sparsedense}.
However, during retrieving candidates, the documents can be passages in natural language~\cite{nguyen2016ms,kwiatkowski2019natural}, images~\cite{chen2015microsoft}, structured data documents~\cite{DBLP:conf/nips/LuGRHSBCDJTLZSZ21} or multi-modal documents~\cite{chang2021webqa}, which challenges existing dense retrieval models to handle different kinds of modalities of knowledge sources to build a self-contained retrieval system. 

Existing work~\cite{guo2020graphcodebert} also builds dense retrievers for retrieving structured data and mainly focuses on learning representations for code data. Leaning more effective representations with PLMs is crucial for dense retrieval~\cite{cocondenser,luan2020sparsedense}, thus several continuous training models are proposed. They usually employ mask language modeling to train PLMs on structured data and help to memorize the semantic knowledge using model parameters~\cite{wang2021codet5,feng2020codebert,roziere2021dobf}. 
%During language model pretraining, the mask strategies are critical~\cite{liu2019roberta,wettig2022should}. These methods target masking and predicting important tokens~\cite{gu2020train}, mentioned references~\cite{ye2020coreferential} or entities~\cite{zhang2019ernie}, making pretraining strategies more effective. 

CodeBERT uses replaced token detection~\cite{clark2020electra} and masked language modeling~\cite{devlin2019bert} to learn the lexical semantics of structured data~\cite{DBLP:conf/nips/LuGRHSBCDJTLZSZ21}. DOBF~\cite{roziere2021dobf} further considers the characteristics of code-related tasks and replaces class, function and variable names with special tokens.
CodeT5~\cite{wang2021codet5} not only employs the span mask strategy~\cite{raffel2020exploring} but also masks the identifiers in codes to teach T5~\cite{raffel2020exploring} to generate these identifiers, which helps better distinguish and comprehend the identifier information in code-related tasks. Nevertheless, the mask language modeling~\cite{devlin2019bert} may not sufficiently train PLMs to represent texts and show less effectiveness in text matching tasks~\cite{chen2021exploring,gao2019representation,li2020sentence,reimers2019sentence,li2020sentence}.

%During language model pretraining, the mask strategies are critical~\cite{liu2019roberta,wettig2022should} to memorize the language knowledge using model parameters. These methods target masking and predicting important tokens~\cite{gu2020train}, mentioned references~\cite{ye2020coreferential} or entities~\cite{zhang2019ernie}, making pretraining strategies more effective. Moreover, 

The recent development of sentence representation learning methods has achieved convincing results~\cite{fang2020cert,yan2021consert}. The work first constructs sentence pairs using back-translation~\cite{fang2020cert}, some easy deformation operations~\cite{wu2020clear}, original sequence cropping~\cite{meng2021coco} or adding dropout noise~\cite{gao2021simcse}. Then they contrastively train PLMs to learn sentence representations that can be used to distinguish the matched sentence pairs with similar semantics.

\begin{figure*}[ht]
\centering
\includegraphics[scale=0.4]{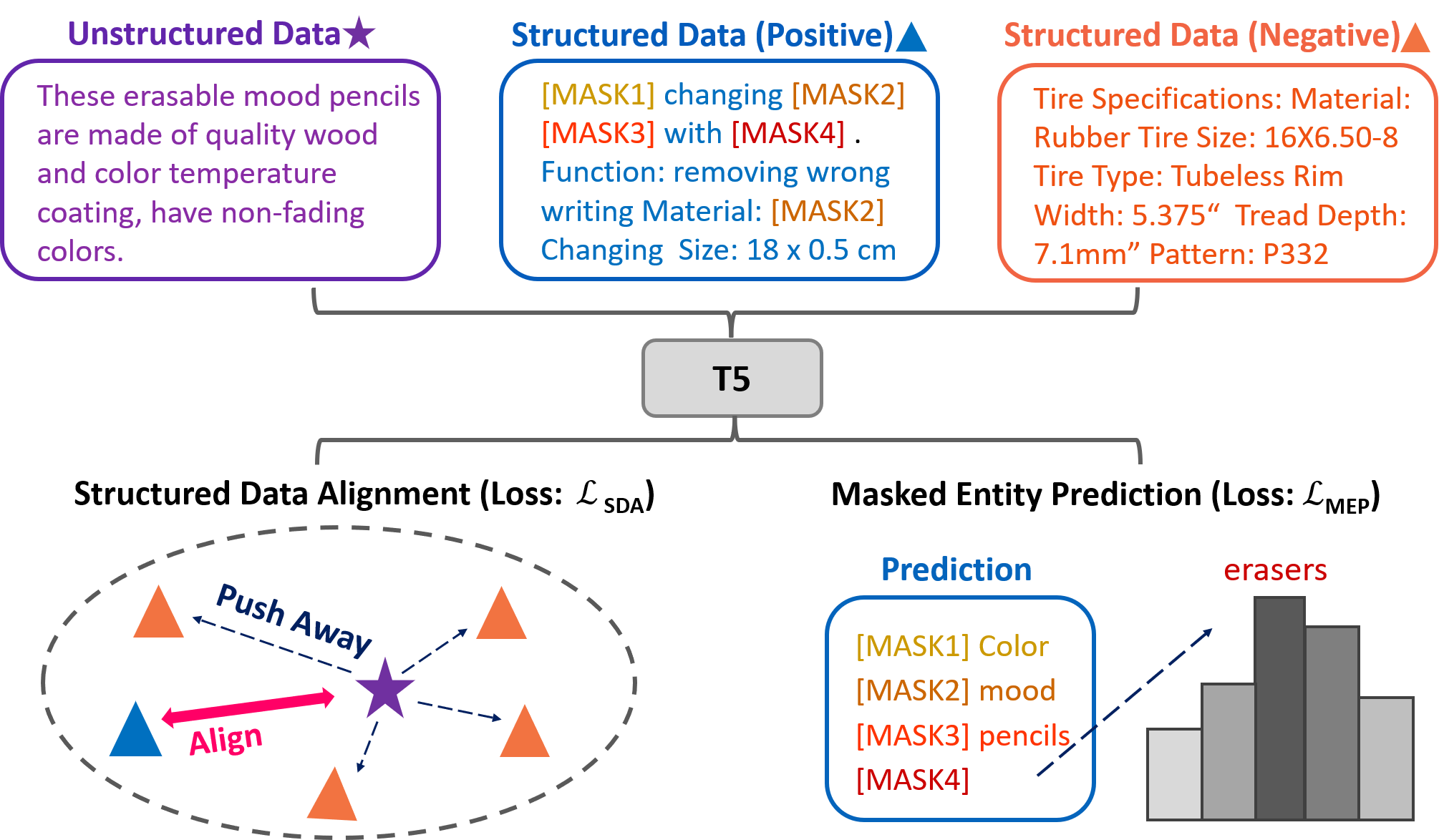}
\caption{The Structure-Aware Pretraining Methods of SANTA. We use both Structured Data Alignment (SDA) and Masked Entity Prediction (MEP) methods for pretraining.}
\label{fig:model}
\end{figure*}
%For SDA task, we use structured data that has been masked entities to align unstructured data.

\section{Methodology}
In this section, we introduce our \textbf{S}tructure \textbf{A}ware De\textbf{N}se Re\textbf{T}riev\textbf{A}l (SANTA) model. First, we introduce the preliminary of dense retrieval (Sec.~\ref{sec:dense}). And then we describe our structure-aware pretraining method (Sec.~\ref{sec:training}).

\subsection{Preliminary of Dense Retrieval}\label{sec:dense} 
Given a query $q$ and a structured data document $d$, dense retriever~\cite{karpukhin2020dense,xiong2020approximate} encodes queries and structured data documents with pretrained language models~\cite{devlin2019bert,liu2019roberta} and maps them in an embedding space for retrieval.

Following previous work~\cite{sentencet5}, we can use T5~\cite{raffel2020exploring} to encode the query $q$ and structured data document $d$ as low dimensional representations $h_q$ and $h_d$, using the representation of the first token from the decoder:
\begin{equation}
\small
h_{q} = \text{T5}(q); h_{d} = \text{T5}(d).
\end{equation}
Then we can calculate the similarity score $f(q, d)$ between the representations of query $h_q$ and structured data document $h_d$: 
\begin{equation}
\small
f(q,d) = sim(h_{q},h_{d}),
\end{equation}
where $sim$ is the dot product function to calculate the relevance between query $q$ and structured data document $d$. %In our method, it is used as the similarity function.

Finally, we can finetune the representations of query and document by minimizing the loss $\mathcal{L}_{\text{DR}}$:
\begin{equation}
\small
 \mathcal{L}_{\text{DR}} = -\log\frac{e^{f(q,d^+)}}{e^{f(q,d^+)} + \sum_{d^-\in \mathcal{D}^-}{e^{f(q,d^-)}}},
\end{equation}
where $d^+$ is relevant to the given query $q$. $\mathcal{D}^-$ is the collection of irrelevant structured data documents, which are sampled from inbatch negatives~\cite{karpukhin2020dense} or hard negatives~\cite{xiong2020approximate}.

\subsection{Structure Aware Pretraining}\label{sec:training}
Existing language models are usually pretrained on unstructured natural languages with masked language modeling~\cite{devlin2019bert,liu2019roberta}. Nevertheless, these models struggle to better understand the semantics represented by data structures, which limits the effectiveness of language models in representing structured data for retrieval~\cite{feng2020codebert,wang2021codet5}.

To get more effective representations for structured data, we come up with structure-aware pretraining methods, aiming to help language models better capture the semantics behind the text structures. As shown in Figure~\ref{fig:model}, we continuously finetune T5 using two pretraining tasks by minimizing the following loss function $\mathcal{L}$:
\begin{equation}
\small
\mathcal{L} = \mathcal{L}_{\text{SDA}} + \mathcal{L}_{\text{MEP}},
\end{equation}
where $\mathcal{L}_{\text{SDA}}$ and $\mathcal{L}_{\text{MEP}}$ are two loss functions from structured data alignment (SDA) (Sec.~\ref{sec:sta}) and masked entity prediction (MEP) (Sec.~\ref{sec:mep}), which are two subtasks of our structure-aware language model pretraining method.

\subsubsection{Structured Data Alignment} 
\label{sec:sta}
The structured data alignment task teaches language models to optimize the embedding space by aligning structured data with unstructured data.

For the structured data document $d$, there are usually some natural language passages that share the same semantics with $d$, \textit{e.g.} the descriptions of codes and bullet points of products. With the help of these text passages $p$ in natural language, we can enhance the model's ability in representing structured data by continuously training language models to align the semantics between structured and unstructured data. Through text data alignment, the representations of structured data are benefited from the intrinsic natural language knowledge of pretrained language models.

Specifically, we can use T5 to encode the text passage and structured data document as $h_p$ and $h_d$, respectively, calculate the similarity score $f(p,d)$ between text passage $p$ and structured data document $d$, and then continuously train language models using the contrastive loss $\mathcal{L}_{\text{SDA}}$:
\begin{equation}\label{eq:sta}
\small
\begin{aligned}
 &\mathcal{L}_{\text{SDA}} = -\log \frac{e^{f(p,d^+)}}{e^{f(p,d^+)}+\sum_{d^- \in D^-} e^{f(p,d^-)}}\\
 &= -f(p,d^+) + \log (e^{f(p,d^+)}+\sum_{d^- \in D^-} e^{f(p,d^-)}),
\end{aligned}
\end{equation}
where $D^-$ consists of the irrelevant structured data sampled from in-batch negatives.

As shown in Eq.~\ref{eq:sta}, the structured data alignment training task helps to optimize the pretrained language models to assign similar embedding features to $<p, d^+>$ pairs and pull $d^-$ away from $p$ in the embedding space~\cite{wang2020understanding}. Such a contrastive training method can bridge the semantic gap between structured and unstructured data and map them in one universal embedding space, benefiting learning representations of multi-modal text data~\cite{liu2022universal}.

\subsubsection{Masked Entity Prediction}\label{sec:mep}
The masked entity prediction guides the language models to better understand the semantics of structured data by recovering masked entities.
SANTA masks entities for continuous training instead of using the random masking in mask language modeling~\cite{devlin2019bert,raffel2020exploring}.

As shown in previous work~\cite{sciavolino2021simple,zhang2019ernie}, entity semantics show strong effectiveness in learning text data representations during retrieval. Thus, we first recognize mentioned entities that appeared in the structured data document $X_d=\{x_1, \text{ent}_1, x_2, \text{ent}_2, ..., \text{ent}_n\}$ and mask them as the input for T5 encoder module:
\begin{equation}
\small
X_d^\text{mask} = \{x_1, \text{<mask>}_1, x_2, \text{<mask>}_2, ..., x_n\},
\end{equation}
where $\text{<mask>}_i$ is a special token to denote the $i$-th masked span. We replace the same entity with the same special token. Then 
 we continuously train T5 to recover these masked entities using the following loss function:
\begin{equation}\label{eq:mep}
\small
\mathcal{L}_{\text{MEP}} = \sum_{j=1}^k -\log P(Y_d (t_j)| X_d^\text{mask}, Y_d (t_{1,...,j-1})),
\end{equation}
where $Y_d (t_j)$ denotes the $j$-th token in the sequence $Y_d$. And $Y_d = \{ \text{<mask>}_1, \text{ent}_1, ..., \text{<mask>}_n, \text{ent}_n\}$ denotes the ground truth sequence that contains masked entities. During training, we optimize the language model to fill up masked spans and better capture entity semantics by picking up the necessary information from contexts to recover the masked entities, understanding the structure semantics of text data, and aligning coherent entities in the structured data~\cite{ye2020coreferential}.

%masked language modeling (MLM) pre-training has been shown to be quite effective in a broad set of NLP tasks. For BERT, they randomly sample 15\% of the tokens from the input sequence, and then replace them with a [MASK] token. Different from the random masking in BERT , we mask all entities in the structured data and mask all occurrences of this entity with the same mask special token. For example, in Figure, we will replace the two occurrences of same entity by the same mask special token, while MLM will only mask one of these occurrences. We name this task to Masked Entity Prediction (MEP), as illustrated in Figure. 

%Entities in code data such as function names, variable names and class names etc. can be easily found using the tokenize tool. But the entities information in the product description is not so easy to define like code. First, we use ultk tool to find special nouns in product title, which contain important information such as product names, attributes, and more. If these nouns exist in product description, we think of them as entities and mask them. In addition, there are some special nouns in the product description, they do not appear in the title, but they also contain some key information of the product, and we treat them as entities.

\section{Experimental Methodology}
In this section, we describe the datasets, evaluation metrics, baselines, and implementation details in our experiments.

\textbf{Dataset.} The datasets in our experiments consist of two parts, which are used for continuous training and finetuning, respectively.

\textit{Continuous Training.}
During continuous training, two datasets, CodeSearchNet~\cite{DBLP:journals/corr/abs-1909-09436} and ESCI (large)~\cite{DBLP:journals/corr/abs-2206-06588}, are employed to continuously train PLMs to conduct structure-aware text representations for codes and shopping products. In our experiments, we regard code documentation descriptions and product bullet points as unstructured data for aligning structured data, codes and product descriptions, during training. More details of pretraining data processing are shown in Appendix~\ref{app:Pretraining Data}.
%The code documentation describes code usages in natural language. And the bullet points and descriptions of products are the key point texts and structured text documents, respectively.

\textit{Finetuning.} For downstream retrieval tasks on structured data, we use Adv~\cite{DBLP:conf/nips/LuGRHSBCDJTLZSZ21}, and ESCI (small)~\cite{DBLP:journals/corr/abs-2206-06588} to finetune models for code search and product search, respectively.
All data statistics are shown in Table~\ref{tab:dataset}. Each query in ESCI (small) has 20 products on average, which are annotated with four-class relevance labels: Exact, Substitute, Complement, and Irrelevant. We also establish a two-class testing scenario by only regarding the products that are annotated with the Exact label as relevant ones. 
%In addition, we also evaluate models on different kinds of code data using the CodeSearch dataset. We provide a more detailed description on CodeSearch dataset in Appendix.~\ref{app:CSN}.

\begin{table}
\begin{center}
\small
\begin{tabular}{l|r|r r}
\hline
\multirow{2}{*}{\textbf{Split}} & \multicolumn{1}{c}{\textbf{Code Search}} & \multicolumn{2}{c}{\textbf{Product Search}} \\ \cline{2-4}
& Query-Code Pair & Query & Product\\
\hline
Train & 251,820 & 18,277 & 367,946 \\
Dev & 9,604 & 2,611 & 51,706 \\
Test & 19,210 & 8,956 & 181,701\\ 
\hline
\end{tabular}
\caption{Data Statistics of Model Finetuning.}
\label{tab:dataset}
\end{center}
\end{table}

\textbf{Evaluation Metrics.} We use MRR@100 and NDCG@100 to evaluate model performance, which is the same as the previous work~\cite{DBLP:conf/nips/LuGRHSBCDJTLZSZ21,DBLP:journals/corr/abs-2206-06588,feng2020codebert}.

\textbf{Baselines.}
We compare SANTA with several dense retrieval models on code search and product search tasks.

We first employ three pretrained language models to build dense retrievers for structured data retrieval, including BERT~\cite{devlin2019bert}, RoBERTa~\cite{liu2019roberta} and T5~\cite{raffel2020exploring}, which are widely used in existing dense retrieval models~\cite{karpukhin2020dense,xiong2020approximate,sentencet5}. All these models are trained with in-batch negatives~\cite{karpukhin2020dense}.%All baseline models use dual encoding for structured data dense retrieval tasks.

For the code search task, we also compare SANTA with three typical and task-specific models, CodeBERT~\cite{feng2020codebert}, CodeT5~\cite{wang2021codet5} and CodeRetriever~\cite{li2022coderetriever}. CodeBERT inherits the BERT architecture and is trained on code corpus using both mask language modeling and replaced token detection. CodeT5 employs the encoder-decoder architecture for modeling different code-related tasks and teaches the model to focus more on code identifiers. CodeRetriever is the state-of-the-art, which continuously trains GraphCodeBERT~\cite{guo2020graphcodebert} with unimodal and bimodal contrastive training losses.

\textbf{Implementation Details.}
This part describes the experiment details of SANTA.

%\textit{Code retriever}. In this task, Our model is initialized with CodeT5. Following previous work, we use code tokens from the dataset as input. In Adv and CodeSearchNet dataset, code tokens convert code into a sequence, removing comments and line breaks from the code. 

%\textit{Product retriever}. In this task, Our model is initialized with T5. Each product in the esci-data has a title, but some product lack a description. So, we concatenate the product title and description as the product information to retrieve. If a product lack description, we just use the title as the product information. There are many HTML symbols in the product description, which are noise information, so we chose to delete them.

We initialize SANTA with T5-base and CodeT5-base for product search and code search. For masked entity prediction, we regard code identifiers and some noun phrases as entities in codes and product descriptions, respectively. More details about identifying entities are shown in Appendix~\ref{app:entity}.

%During continuous training, we set the training epoch to 60 and learning rate to 1e-4 for product search, while the training epoch and learning rate are 6 and 5e-5 for code search. 
%The learning rates are set to 5e-5 and 1e-5 for product search and code search during training SANTA using inbatch negatives. 
During continuous training, we set the learning rate as 1e-4 and 5e-5 for product search and code search, and the training epoch as 6. During finetuning, we conduct experiments by training SANTA using inbatch negatives and hard negatives. we set the training epoch to 60 and learning rate to 5e-5 for product search, while the training epoch and learning rate are 6 and 1e-5 for code search. And we follow ANCE~\cite{xiong2020approximate}, start from inbatch finetuned SANTA (Inbatch) model and continuously finetune it with hard negatives to conduct the SANTA (Hard Negative) model. The learning rates are set to 1e-5 and 1e-6 for product search and code search. These hard negatives are randomly sampled from the top 100 retrieved negative codes/product descriptions from the SANTA (Inbatch) model.

All models are implemented with PyTorch, Huggingface transformers~\cite{wolf2019huggingface} and OpenMatch~\cite{liu2021openmatch}. We use Adam optimizer to optimize SANTA, set the batch size to 16 and set the warmup proportion to 0.1 in our experiments.

\section{Evaluation Results}
In this section, we focus on exploring the performance of SANTA on code search and product search tasks, the advantages of SANTA in representing structured data, and the effectiveness of proposed pretraining methods.
\begin{table}
\begin{center}
\resizebox{0.49\textwidth}{!}{
\begin{tabular}{l | c  c c}
\hline
\multirow{3}{*}{\textbf{Model}} & \textbf{Code} &  \multicolumn{2}{c}{\textbf{Product}} \\ \cline{2-4}
& \multirow{2}{*}{MRR} & \multicolumn{2}{c}{NDCG} \\ \cline{3-4}
& & Two-C & Four-C \\
\hline
\multicolumn{2}{l}{\textit{\textbf{Zero-Shot}}} \\\hline
BERT~\cite{devlin2019bert} & 0.20 & 71.46 & 72.45\\
RoBERTa~\cite{liu2019roberta} & 0.03 & 71.25 & 72.24 \\
CodeBERT~\cite{feng2020codebert} & 0.03 & - & - \\
CodeRetriever~\cite{li2022coderetriever} & 34.7 & - & -\\
T5~\cite{raffel2020exploring} & 0.03 & 70.21 & 71.25 \\
CodeT5~\cite{wang2021codet5} & 0.03 & - & -\\
SANTA  & \textbf{46.1} & \textbf{76.38} & \textbf{77.14} \\
\hline
\multicolumn{2}{l}{\textit{\textbf{Fine-Tuning}}} \\\hline
BERT~\cite{devlin2019bert} & 16.7 & 78.29 & 79.06\\
RoBERTa~\cite{liu2019roberta} & 18.3 & 79.59 & 80.29 \\
CodeBERT~\cite{feng2020codebert} & 27.2 & - & - \\
CodeRetriever & 43.0 & - & -\\
%CodeRetriever (Hard Negative) & 45.1 & - & -\\
CodeRetriever (AR2)~\cite{li2022coderetriever} & 46.9 & - & -\\
T5~\cite{raffel2020exploring} & 23.8 & 79.77 & 80.46 \\
CodeT5~\cite{wang2021codet5} & 39.3 & - & -\\
SANTA (Inbatch)  & 47.3 & 80.76 & 81.41 \\
SANTA (Hard Negative) & \textbf{47.5} & \textbf{82.59} & \textbf{83.15}\\
\hline
\end{tabular}}
\caption{Retrieval Effectiveness of Different Models on Structured Data. For product search, there are two ways to evaluate model performance. Two-C regards the query-product relevance as two classes, Relevant (1) and Irrelevant (0). Four-C is consistent with the ESCI dataset~\cite{DBLP:journals/corr/abs-2206-06588} and sets the relevance labels with the following four classes: Exact (1), Substitute (0.1), Complement (0.01), and Irrelevant (0).}
\label{tab:codesearch}
\end{center}
\end{table}

\subsection{Overall Performance}
The performance of SANTA on structured data retrieval is shown in Table~\ref{tab:codesearch}.

SANTA shows strong zero-shot ability by comparing its performance with finetuned models and achieving 6.8\% improvements over finetuned CodeT5 on code search. Such impressive improvements demonstrate that our pretrained strategies have the ability to enable the advantages of PLMs in representing structured data without finetuning.

After finetuning, SANTA maintains its advantages by achieving about 8\% and 2\% improvements over CodeT5 and T5 on code search and product search, respectively. It shows the critical role of structure-aware pretraining, which makes language models sensitive to text data structures and better represents structured data. On code retrieval, SANTA outperforms the state-of-the-art code retrieval model CodeRetriever with 4.3\% improvements under the same inbatch training setting. SANTA also beats CodeRetriever (AR2), which is finetuned with more sophisticated training strategies~\cite{ZhangGS0DC22} and the larger batch size. 

Besides, we show the retrieval performance of SANTA on CodeSearch dataset in Appendix~\ref{app:CSN}.

\begin{table}
\centering
\small
\begin{tabular}{l | c | c c}

\hline
\multirow{3}{*}{\textbf{Model}} & \textbf{Code} & \multicolumn{2}{c}{\textbf{Product}}\\ \cline{2-4}
& \multirow{2}{*}{MRR} & \multicolumn{2}{c}{NDCG} \\ \cline{3-4}
& & Two-C & Four-C \\
\hline
\multicolumn{2}{l}{\textit{\textbf{Zero-Shot}}} \\\hline
T5 (Baseline) & 0.03 & 70.21 & 71.25  \\
T5 (w/ MEP) & 0.03 & 70.56 & 71.58 \\
T5 (w/ SDA) & 45.01 & 76.64 & 77.40\\
SANTA  (Span Mask) & 35.88 & \textbf{77.37} & \textbf{78.11} \\
SANTA  (Entity Mask)& \textbf{46.08} & 76.38 & 77.14 \\
\hline
\multicolumn{2}{l}{\textit{\textbf{Fine-Tuning}}} \\\hline
T5 (Baseline) & 39.30 & 79.77 & 80.46 \\
T5 (w/ MEP) & 38.46  & 79.50 & 80.29 \\
T5 (w/ SDA) & 46.98 & 80.42 & 81.11\\
SANTA  (Span Mask) & 42.11 & 80.31 & 80.99 \\
SANTA  (Entity Mask) & \textbf{47.28} & \textbf{80.76} & \textbf{81.41} \\
\hline
\end{tabular}
\caption{The Retrieval Performance of Ablation Models of SANTA on Structured Data Retrieval. Masked Entity Prediction (MEP) and Structured Data Alignment (SDA) are two pretrained tasks that are proposed by SANTA.}
\label{tab:ablation}
\end{table}

\subsection{Ablation Study}
In this subsection, we conduct ablation studies to further explore the
roles of different components in SANTA on retrieving structured data.

We start from CodeT5/T5 models and continuously train CodeT5/T5 using two proposed training tasks, Masked Entity Prediction (MEP) and Structured Data Alignment (SDA) to show their effectiveness in teaching models to better learn semantics from structured data. Meanwhile, we compare MEP with the random span masking strategy~\cite{raffel2020exploring,wang2021codet5} to evaluate the effectiveness of different masking strategies. The retrieval performance in both zero-shot and finetuning settings is shown in Table~\ref{tab:ablation}.

Compared with our baseline model, MEP and SDA show distinct performance in structured data retrieval. As expected, MEP shows almost the same performance as the baseline model. It shows that only mask language modeling usually shows less effectiveness in learning representations for structured data, even using different masking strategies.
Different from MEP, SDA shows significant improvements in both structured data retrieval tasks, especially the code retrieval task. Our SDA training method contrastively trains T5 models using the alignment relations between structured data and unstructured data, which helps to bridge the modality gap between structured and unstructured data, maps structured and unstructured data in one universal embedding space, and learns more effective representations for retrieval. When adding additional task MEP to T5 (w/ SDA), the retrieval performance of SANTA is consistently improved. This phenomenon shows that mask language modeling is still effective to teach T5 to better capture the structure semantics and conduct more effective text representations for structured data by filling up the masked entities of structured data.

We also compare different masking strategies that are used during mask language modeling. Our entity masking strategy usually outperforms the random span masking strategy, showing the crucial role of entities in structured data understanding. With the masked entity prediction task, SANTA achieves comparable ranking performance with finetuned models, which illustrates that structure-aware pretraining is starting to benefit downstream tasks, such as structured data retrieval. The next experiment further explores how these pretraining strategies guide models to learn representations of structured/unstructured data.

\subsection{Embedding Visualization of Structured and Unstructured Data}
This section further explores the characteristics of embedding distributions of structured and unstructured data learned by SANTA.

\begin{figure}[t]
    \centering 
    \subfigure[Ranking Probability of Matched Text Data Pairs.] { \label{fig:retrieval:prob} 
\includegraphics[width=0.48\linewidth]{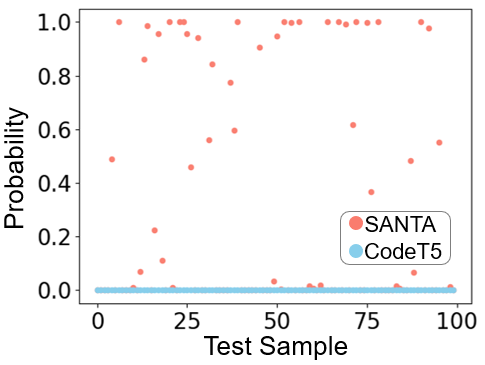}}
     \subfigure[Embedding Distribution of Structured Data.] { \label{fig:retrieval:embed} 
    \includegraphics[width=0.48\linewidth]{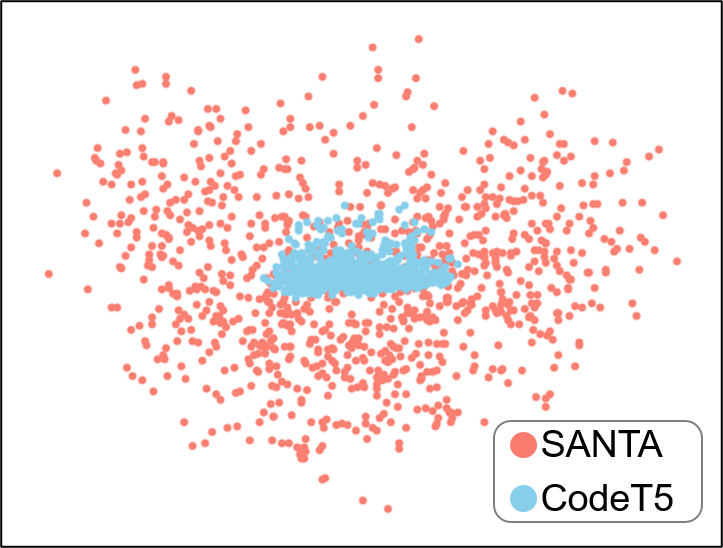}}
    \caption{Retrieval Effectiveness on Code Search. We sample several query-code pairs from the test split of code search data and show the ranking probability distribution of query-related codes in Figure~\ref{fig:retrieval:prob}. Then Figure~\ref{fig:retrieval:embed} presents the learned embedding space of structured data of codes.}
    \label{fig:retrieval}
\end{figure}

As shown in Figure~\ref{fig:retrieval}, we first conduct experiments to show the retrieval effectiveness of CodeT5 and SANTA under the zero-shot setting. The ranking probability distribution of relevant query-code pairs is shown in Figure~\ref{fig:retrieval:prob}. Even though CodeT5 is pretrained with code text data, it seems that CodeT5 learns ineffective representations for structured data, assigns a uniform ranking probability distribution for all testing examples and fails to pick up the related structured data for the given queries. On the contrary, SANTA assigns much higher ranking probabilities to matched structured documents, demonstrating that our structured data alignment task has the ability to guide the model to conduct more effective text data representations to align queries with its relevant structured documents. Then we plot the embedding distribution of structured data in Figure~\ref{fig:retrieval:embed}. Distinct from the embedding distribution of CodeT5, the embeddings learned by SANTA, are more distinguishable and uniform, which are two criteria of learning more effective embedding space under contrastive training~\cite{li2021more,wang2020understanding}.

\begin{figure}[t]
    \centering
    \subfigure[CodeT5.] { \label{fig:embed:codet5} 
    \includegraphics[width=0.48\linewidth]{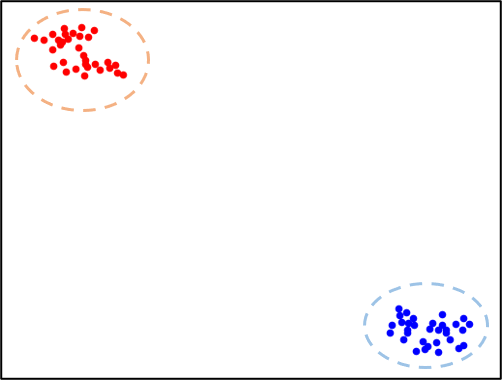}}
    \subfigure[CodeT5 (w/ SDA).] { \label{fig:embed:CL} 
    \includegraphics[width=0.48\linewidth]{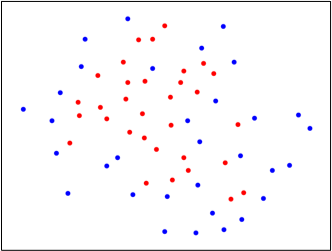}}

    \subfigure[CodeT5 (w/ MEP).] { \label{fig:embed:mlm} 
    \includegraphics[width=0.48\linewidth]{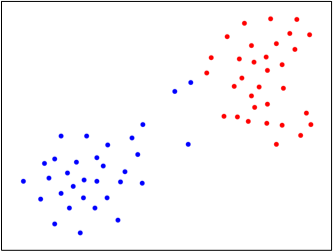}}
    \subfigure[SANTA.] { \label{fig:embed:SANTA}
    \includegraphics[width=0.48\linewidth]{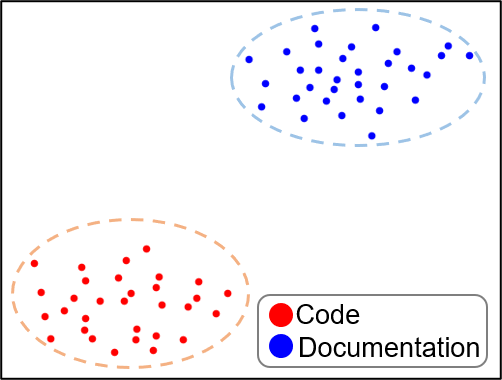}}
    \caption{Embedding Visualization of Different Models using T-SNE. We randomly sample 32 codes and 32 code documentation texts from the testing set of code retrieval and plot their embedding distribution.}
    \label{fig:embed}
\end{figure}

Then we present the embedding distribution of documentation texts and their corresponding codes in Figure~\ref{fig:embed}. Overall, depending on our structure-aware pretraining methods, SANTA conducts a more uniform embedding space than CodeT5 and makes the representations of structured and unstructured data more distinguished in the embedding space. Then we analyze the effectiveness of our continuous training methods, Masked Entity Prediction (MEP) and Structured Data Alignment (SDA).
By comparing Figure~\ref{fig:embed:CL} with Figure~\ref{fig:embed:codet5}, our structured data alignment task indeed helps PLMs to align the representations of code and documentation, which reduces the distance between matched unstructured-structured data pairs and mixes the multi-modal embeddings thoroughly in the embedding space. After adding the masked entity prediction training task to CodeT5 (w/ SDA) (from Figure~\ref{fig:embed:CL} to Figure~\ref{fig:embed:SANTA}), the embedding distributions of code and documentation become distinguished again, demonstrating that masked entity prediction can help models capture different semantics from different data modalities to represent unstructured/structured data. Besides, by comparing Figure~\ref{fig:embed:SANTA} with Figure~\ref{fig:embed:mlm}, the structured data alignment task also makes the boundary of the embedding clusters of code and documentation clearer. The main reason lies in that these embeddings are assigned to appropriate positions for aligning matched code-documentation pairs with the help of our structured data alignment task.

%In this experiment, we visualize the data embedding space in Figure~\ref{fig:case}. By comparing picture (a) and (b), we can find that after using structured and unstructured text alignment tasks, the types of text are no longer independently distributed in vector space, but mixed together, narrowing the distance between structured and unstructured text. But in Figure (d) we find that after training with multitask, the distribution of unstructured text and structured text is independent again.

\begin{table*}[!ht]
\small
\begin{tabular}{p{0.1\linewidth}|p{0.4\linewidth}|p{0.4\linewidth}}
\hline 
\textbf{Model}  & \textbf{ SANTA}& \textbf{CodeT5/T5} 
\tabularnewline 
\hline
\textbf{Query} & \multicolumn{2}{p{0.8\linewidth}}{Construct the command to \textcolor[rgb]{0.7,0.3,0.3}{\textbf{poll the driver status}}}
\tabularnewline 
\hline
Rank & 1 & 1
\tabularnewline 
\hline
Snippet 
&
 ... arg\_0 . \_connection [ 'master' ] ] if arg\_0 . \_driver\_id : arg\_1 += [ "--\textcolor[rgb]{0.7,0.3,0.3}{\textbf{status}}" , arg\_0 . \_driver\_id ] else : raise AirflowException ( "--\textcolor[rgb]{0.7,0.3,0.3}{\textbf{Invalid status: attempted to poll driver}} ...
&
def Func ( arg\_0 ) : return os . path . join ( get\_user\_config\_dir ( arg\_0 . app\_name , arg\_0 . app\_author ) , arg\_0 . filename )

\tabularnewline 
\hline
\textbf{Query} & \multicolumn{2}{p{0.8\linewidth}}{Attempt to \textcolor[rgb]{0.7,0.3,0.3}{\textbf{copy path with storage}}.}
\tabularnewline 
\hline
Rank &  1  &  1
\tabularnewline 
\hline
Snippet 
&
 ... if arg\_2 in arg\_0 . copied\_files : return arg\_0 . log ( "Skipping '\%s' (already  --\textcolor[rgb]{0.7,0.3,0.3}{\textbf{copied}} earlier)" \% arg\_1 ) if not arg\_0 . delete\_file ( arg\_1 , arg\_2 , arg\_3 ) : return arg\_4 = arg\_3 . --\textcolor[rgb]{0.7,0.3,0.3}{\textbf{path }} ( arg\_1 ) ...
&
'... arg\_0 ) : if arg\_0 . \_api\_arg : arg\_1 = str ( arg\_0 .\_api\_arg ) else : arg\_1 = arg\_0 . \_name if arg\_0 . \_parent : return '/' . join ( filter ( None , [ arg\_0 . \_parent . Func , arg\_1 ] ) ) ...'
\tabularnewline 
\hline 
\textbf{Query} & \multicolumn{2}{p{0.7\linewidth}}{\#1 black natural \textcolor[rgb]{0.7,0.3,0.3}{\textbf{hair dye without ammonia or peroxide}}}
\tabularnewline 
\hline
Rank & 1  &  1
\tabularnewline 
\hline
Snippet &
... naturcolor Haircolor \textcolor[rgb]{0.7,0.3,0.3}{\textbf{Hair Dye}} - Light Burdock, 4 Ounce (5N) naturcolor 5n light burdock permanent herbal Ingredients: haircolor gel utilizes herbs to cover grey as \textcolor[rgb]{0.7,0.3,0.3}{\textbf{opposed to chemicals}} ...&
... Naturtint Permanent Hair Color 5N Light Chestnut Brown (Pack of 1), Ammonia Free, Vegan, Cruelty Free, up to 100\% Gray Coverage, Long Lasting Results...
\tabularnewline 
\hline 
\textbf{Query} & \multicolumn{2}{p{0.7\linewidth}}{!\textcolor[rgb]{0.7,0.3,0.3}{\textbf{qscreen fence}} without holes}

\tabularnewline 
\hline
Rank & 2 & 2
\tabularnewline 
\hline
Snippet &
... Material: HDPE+Brass Color: Green Size(L x W): About 6’x50” Package included: Garden \textcolor[rgb]{0.7,0.3,0.3}{\textbf{fence privacy screen}}*1 Straps*80 ... &
... Windscreen Cover Fabric Shade Tarp Netting Mesh Cloth - Commercial Grade 170 GSM - Cable Zip Ties Included - We Make Custom Size..
\tabularnewline 
\hline 
\end{tabular}
  \caption{Case Studies. We sample four cases from the test datasets of code search and product search to show the effectiveness of SANTA. The matched text phrases are \textcolor[rgb]{0.7,0.3,0.3}{\textbf{highlighted}}.}
  \label{tab:case_study}
\end{table*}
\begin{figure}[t]
    \centering 
    \subfigure[Code Search.] { \label{fig:codeattention}
     \includegraphics[width=0.99\linewidth]{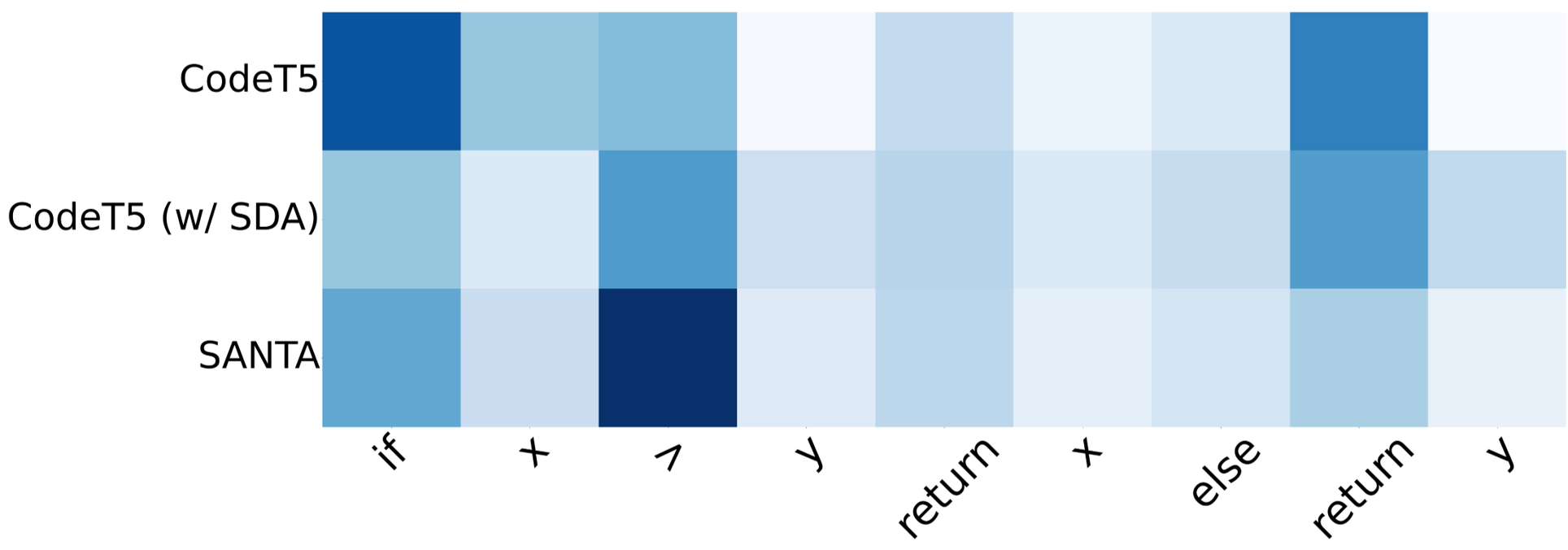}}
     
     \subfigure[Product Search.] { \label{fig:productattention} 
    \includegraphics[width=0.99\linewidth]{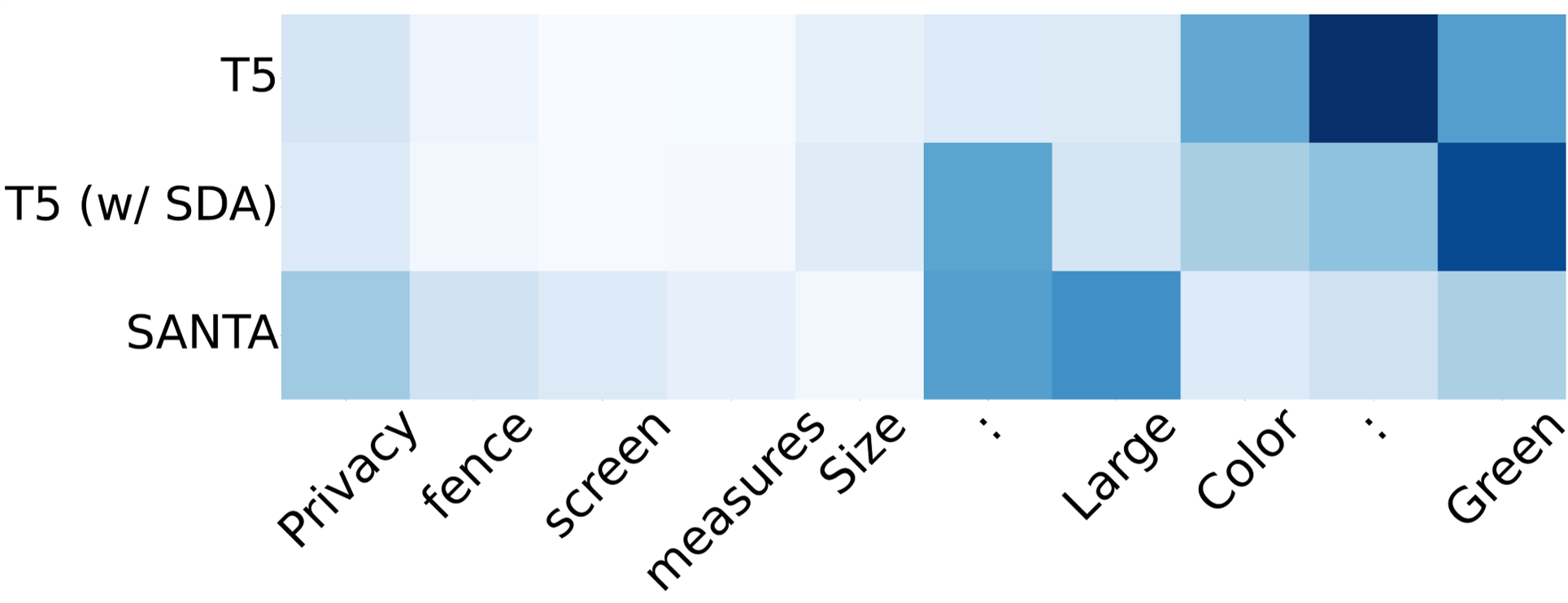}}
    \caption{Visualization of Attention Distribution of SANTA. The cross attention weight distributions from the decoder module to encoded token embeddings are plotted. Darker blue indicates a higher attention weight.} %We randomly sample a small piece of code and a text sequence of product descriptions from code search and product search to plot the attention distribution in Figure~\ref{fig:codeattention} and Figure~\ref{fig:productattention}, respectively. 
    \label{fig:heat}
\end{figure}
\subsection{Attention Mechanism of SANTA}
This section presents the attention mechanism of SANTA during encoding structured data. In Figure~\ref{fig:heat}, we randomly sample a small piece of code and a text sequence of product descriptions to plot the attention distribution.

The attention weight distributions on code search are shown in Figure~\ref{fig:codeattention}. Compared with CodeT5, CodeT5 (w/ SDA) and SANTA calibrate the attention weights from the ``if'' token to the ``>'' token. The ``>'' token is a logical operation, which indicates the usage of the code. SANTA thrives on the structured data alignment task and captures these important semantic clues to represent codes. Compared with CodeT5 (w/ SDA), SANTA decreases its attention weights on code identifiers, such as ``x'' and ``y'', and shares more attention weights to ``If'' and ``>''. These identifiers can be replaced with attribute ones and are less important than these logical operations to understand code semantics. Thus, SANTA adjusts its attention weights to logical tokens to understand structured data, which is benefited from pretraining with the masked entity prediction task.

Figure~\ref{fig:productattention} shows the attention distribution on product search. T5 (w/ SDA) assigns more attention weights to the product attribute ``Green'' than T5, as well as highlights the sequence boundary tokens of product attributes. Nevertheless, for the product ``privacy fence screen'', ``Large'' is a more important attribute than ``Green''. SANTA captures such semantic relevance, which confirms that our masked entity prediction task indeed helps to improve the semantic understanding ability of language models on structured data.

\subsection{Case Studies}
Finally, we show several cases in Table~\ref{tab:case_study} to analyze the ranking effectiveness of SANTA.

In the first case, SANTA directly matches queries and codes through the text snippet ``poll the driver status''. It demonstrates that SANTA has the ability to distinguish the differences between code and documentation and pick up the necessary text clues for matching queries and codes. Then the second case illustrates that SANTA is effective in understanding codes by capturing the structure semantics of codes and matching queries and codes by capturing some keywords in codes, such as ``copied'' and ``path''. The last two cases are from product search and the product description is more like natural language. SANTA also shows its effectiveness on identifying some important entities, such as ``Hair Dye'' and ``fence screen'', to match queries and products. 

%We first compare attention changes of the three models CodeT5, CodeT5 (w/SDA) and SANTA on a code case. This code case is uesed to find the largest number in two numbers, This code is mainly used for code shackles, in which the most important token is the '>', followed by the 'if' judgment, which forms the backbone of the code. For other code tokens, variable names are not unique, for example 'x' can be replaced with 'a' or other variable names. 'Return' is also not required, it can be replaced by 'print' to directly output the largest number, so these code tokens are minor. For CodeT5, it doesn't pay much attention to other tokens than 'if' and 'return' tokens, probably due to the high frequency of python keywords such as if and return in the code, making CodeT5 pay more attention to them during pretrained. CodeT5 noticed some important information in the structure, but not all. For CodeT5 (w/SDA), its attention is more evenly distributed, which may be due to the effect of alignment, but its attention to important tokens is not particularly higher than that of other tokens. For SANTA, we find that its attention is focused on the key information of structured data, and the attention score of if and > is much higher than other tokens.
\section{Conclusion}
%This paper proposes SANTA, which continuously pretrains language models with two tasks, structured data alignment and masked entity prediction. These tasks help PLMs to capture structure-aware clues to represent structured data and map both queries and structured data in one universal embedding space for retrieval. SANTA achieves state-of-the-art on code and product search.
This paper proposes SANTA, which pretrains language models to understand structure semantics of text data and guides language models to map both queries and structured texts in one universal embedding space for retrieval. SANTA designs both structured text alignment and masked entity prediction tasks to continuously train pretrained language models to learn the semantics behind data structures. Our experiments show that SANTA achieves state-of-the-art on code and product search by learning more tailored representations for structured data, capturing semantics from structured data and bridging the modality gap between structured and unstructured data.
\section*{Limitations}
Even though SANTA shows strong effectiveness on learning the representation of structured data, it heavily depends on the alignment signals between structured and unstructured data. Such alignment relations can be witnessed everywhere, but the quality of constructed pairs of structured and unstructured data directly determines the effectiveness of SANTA. Besides, we use the product bullet points and code descriptions as the unstructured data in our experiments, which is designed for specific tasks and limits the model's generalization ability. On the other hand, SANTA mainly focuses on evaluating the structured data understanding ability through text data representation and matching. It is still unclear whether SANTA outperforms baseline models in all downstream tasks, such as code summarization and code generation. 
%In this paper, we have the following limitations. First, we use Adv dataset for code search, which is the most challenging code search dataset out there. However, there is only python code in the Adv, and no other programming language. Although most programming languages have a similar structure, this does not mean that SANTA will perform well on other programming languages. Second, in the product search dataset, many products don't have descriptions, and some product descriptions don't have obvious structural characteristics. Therefore, the improvement of SANTA on the product retrieval task is not as good as that of the code retrieval.
\section*{Acknowledgments}
This work is supported by the Natural Science Foundation of China under Grant No. 62206042, No. 62137001 and No. 62272093, the Fundamental Research Funds for the Central Universities under Grant No. N2216013 and No. N2216017, China Postdoctoral Science Foundation under Grant No. 2022M710022, and National Science and Technology Major Project (J2019-IV-0002-0069).
\bibliography{references}
\bibliographystyle{acl_natbib}

\clearpage
\appendix
\begin{appendix}

\section{Appendix}
\subsection{License}
For all datasets in our experiments, Adv and CodeSearchNet use MIT License, while ESCI uses Apache License 2.0. All of these licenses and agreements allow their data for academic use.

\begin{figure*}[t]
\centering
\includegraphics[scale=0.38]{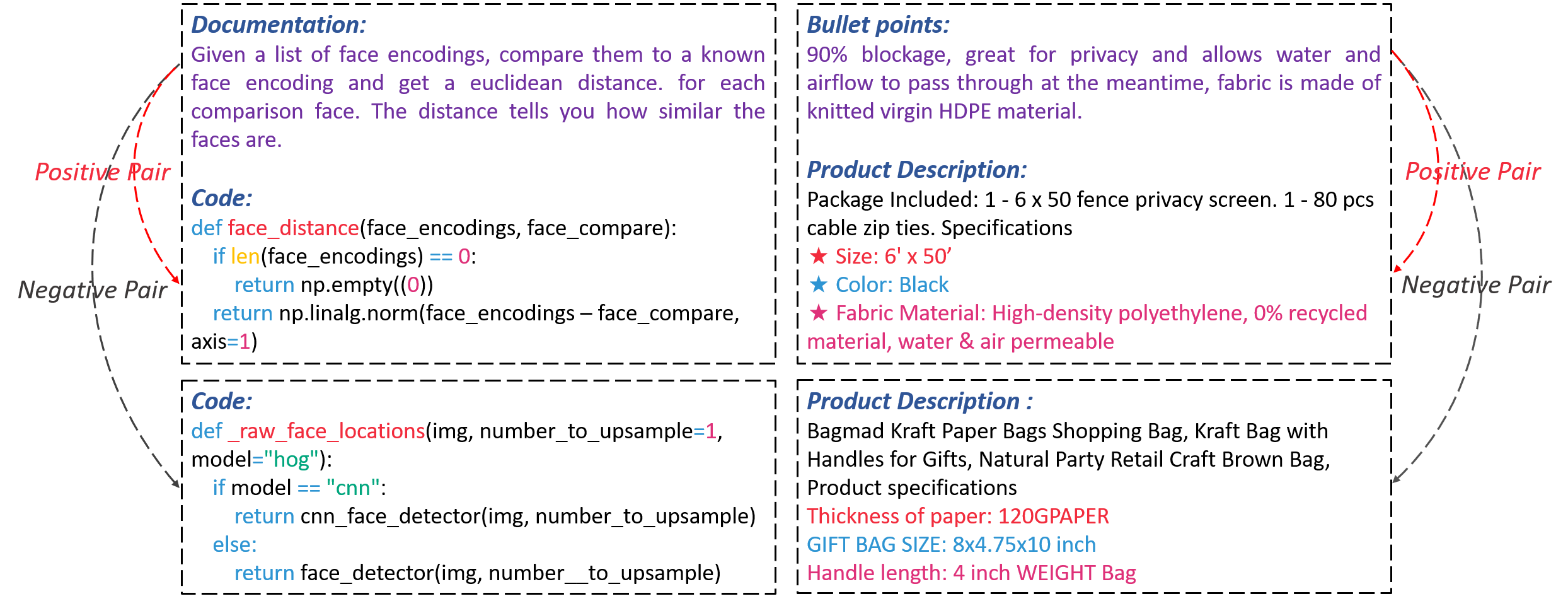}
\caption{Examples of Positive and Negative Pairs of Pretraining Data.}
\label{fig:pretrain data}
\end{figure*}
\begin{table}
\begin{center}
\small

\begin{tabular}{l| r r}
\hline
{\textbf{Task}} & {\textbf{Positive Pairs}} & {\textbf{Entities}} \\ 
\hline
Python & 429,596 & 28.6\%  \\
PHP & 514,127 &  17.8\% \\
Go & 317,824 & 17.1\% \\ 
Java & 454,433 & 24.4\%  \\
JavaScript & 122,682 & 15.4\% \\
Ruby  & 487,90 & 28.8\% \\
\hline
Product & 331,590 & 20.1\% \\
\hline
\end{tabular}
\caption{Data Statistics of Pretraining Data. ``Entities'' denotes the proportion of identified entities in the structured data.}
\label{tab:pretrain-data}
\end{center}
\end{table}
\subsection{Construction of Pretraining Data}\label{app:Pretraining Data}
In this subsection, we show how to process the pretraining data and construct structured-unstructured data for code/product search. During pretraining, we use inbatch negatives to optimize SANTA and all data statistics are shown in Table~\ref{tab:pretrain-data}.

As shown in Figure~\ref{fig:pretrain data}, we show some examples to show how to construct structured-unstructured data pairs for pretraining. For code retrieval tasks, code snippets have corresponding documentation descriptions, which describe the purpose and function of these code snippets. Thus, the code documentation and its corresponding code snippet are regarded as a positive training pair.
%https://github.com/salesforce/CodeT5/issues/64
%\footnote{CodeT5 and CodeRetriever collect 453,772 and 449,216 python data from CodeSearchNet, respectively, both larger than CodeSearchNet's training dataset}

For product retrieval tasks, structured product descriptions usually have corresponding unstructured bullet points, which provide key points about the products. We randomly select one bullet point of items and use its corresponding product description to construct a positive training pair.

\begin{figure}[t]
    \centering 
    \subfigure[Code Search.] { \label{fig:code_entity}
     \includegraphics[width=2.5in]{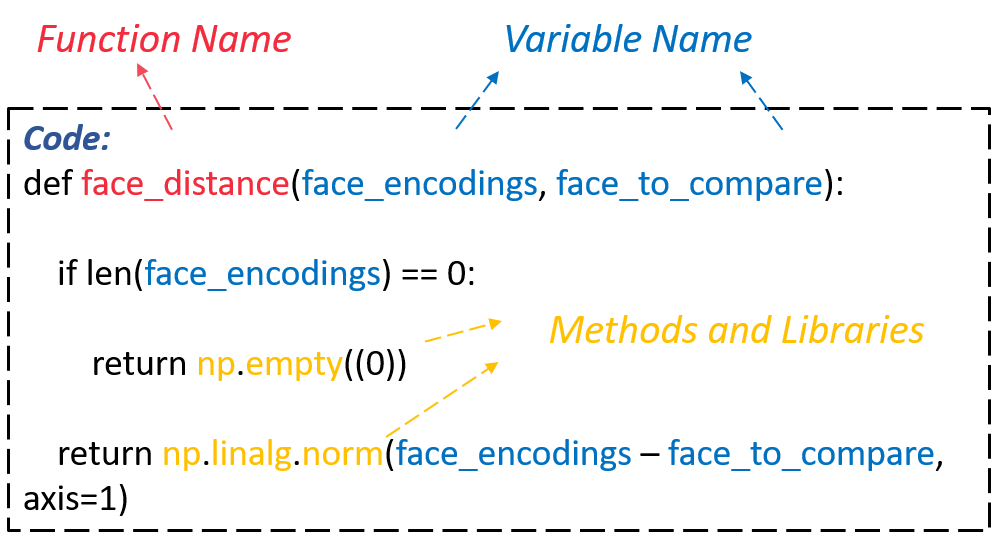}}
     
     \subfigure[Product Search.] { \label{fig:product_entity} 
    \includegraphics[width=2.5in]{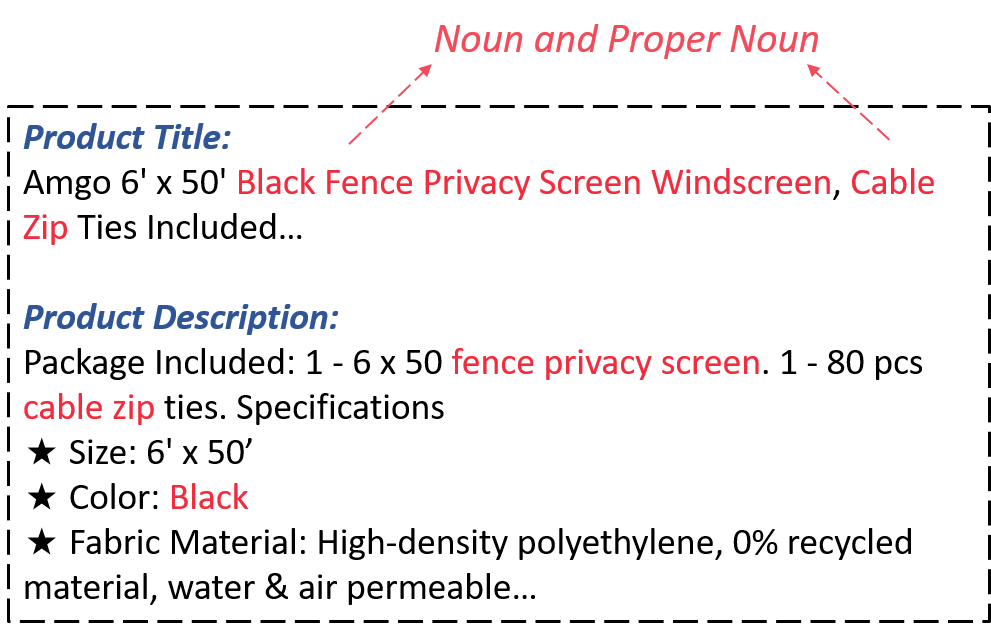}}
    \caption{The Illustration of Identified Entities in Structured Data. All entities of different functions are annotated with different colors.}
    \label{fig:entity}
\end{figure}

\subsection{Additional Experimental Details of Entities Identification on Structured Data}\label{app:entity}
We show some examples of entity identifications on structured data in Figure~\ref{fig:entity}.

For codes, we follow~\citet{wang2021codet5} and regard code identifiers as entities such as variables, function names, external libraries and methods. Specifically, we use BytesIO and tree\_sitter\footnote{\url{https://github.com/tree-sitter/tree-sitter}} to identify entities in Python and other programming languages, respectively. %These tools convert programming languages to abstract syntax trees and obtain their node types. %For codes with comments, we keep more code entity information, because longer and more complex code usually has comments, and keeping these information is more conducive to alignment. Specifically, for some entities that are behind code comments, we mask them if they have already appear before comments, and keep them if they don't exist before comments.
For product descriptions, we use the NLTK tool\footnote{\url{https://www.nltk.org/}} to identify nouns and proper nouns that appear in both product descriptions and titles and regard them as entities. 

In our experiments, we replace the same entities with the same special tokens and ask SANTA to generate these masked entities (Eq.~\ref{eq:mep}). These special tokens come from the predefined vocabulary of T5, such as \{\text{<extra\_id\_0>}, \text{<extra\_id\_1>}, ..., \text{<extra\_id\_99>} \}. The proportions of identified entities in pretraining data are shown in Table~\ref{tab:pretrain-data}.
%In particular, for JavaScript, we only mask about 30\% percent of entities which we selected. Because masking too many entities for JavaScript will affect the training effect. The proportion of entities in different types of structured data is not the same, which is related to the data type and the statistical tools and methods we use.

\begin{table*}[htbp]
\centering
\small
\begin{tabular}{l |c c c c c c c |c}

\hline
\multirow{2}{*}{\textbf{Model}} & \multicolumn{7}{c|}{\textbf{CodeSearch}} & \multirow{2}{*}{\textbf{Adv}} \\  \cline{2-8} & \textbf{Ruby} & \textbf{Javascript} & \textbf{Go}& \textbf{Python}& \textbf{Java}& \textbf{PHP}& \textbf{Overall}\\ 
\hline
\multicolumn{2}{l}{\textit{\textbf{Zero-Shot}}} \\
\hline
GraphCodeBERT & 1.5 & 0.4 & 0.2 & 0.4 & 0.7 & 2.1 & 0.88 & 0.5\\
CodeRetriever & 68.7 & \textbf{63.7} & 87.6 & 67.7 & \textbf{69.0} & 62.8 & 69.1 & 34.7\\
SANTA & \textbf{72.6} & 62.4 & \textbf{88.9} & \textbf{70.0} & 68.6 & \textbf{62.8} &  \textbf{70.9} &  \textbf{48.1} \\\hline
\multicolumn{2}{l}{\textit{\textbf{Fine-Tuning}}} \\
\hline
CodeBERT & 67.9 & 62.0 & 88.2 & 67.2 & 67.6 & 62.8 & 69.3  & 27.2\\
GraphCodeBERT & 70.3 & 64.4 & 89.7 & 69.2 & 69.1 & 64.9 & 71.3 & 35.2\\ 
CodeT5 & 71.9 & 65.5 & 88.8 & 69.8 & 68.6 & 64.5 & 71.5 & 39.3\\
CodeRetriever (Inbatch) & \textbf{75.3} & 69.5 & 91.6 & 73.3 & 74.0 & 68.2 & 75.3 & 43.0\\
CodeRetriever (Hard Negative) & 75.1 & \textbf{69.8} & \textbf{92.3} & \textbf{74.0} & \textbf{74.9} & \textbf{69.1} & \textbf{75.9} & 45.1\\

SANTA (Hard Negative) & 74.7 & 68.6 & 91.8 & 73.7 & 73.7 & 68.6 & 75.2 & \textbf{48.6}\\

\hline
\end{tabular}
\caption{Code Retrieval Evaluations of SANTA. Because of the GPU memory limitation, we set the batch size as 128 during pretraining and finetuning, which is different with~\citet{li2022coderetriever}. All models are evaluated using MRR.}
\label{tab:CSN}
\end{table*}

\begin{table}
\begin{center}
\small

\begin{tabular}{l|r|r|r|r}
\hline
\multirow{2}{*}{\textbf{Language}} & \multicolumn{3}{c|}{\textbf{Query}} & \multirow{2}{*}{\textbf{Document}} \\ \cline{2-4} & {\textbf{Train}} &{\textbf{Dev}} &{\textbf{Test}} & \\ 

\hline
Python & 251,820 & 13,914 & 14,918	 & 43,827  \\
PHP & 241,241 & 12,982 & 14,014 & 52,660  \\
Go & 167,288 & 7,325 & 8,122 & 28,120 \\ 
Java & 164,923 & 5,183 & 10,955 & 40,347    \\
JavaScript & 58,025 & 3,885 & 3,291 & 13,981 \\
Ruby & 24,927 & 1,400 & 1,261 & 4,360 \\
\hline
\end{tabular}
\caption{Data Statistics of CodeSearch Dataset. The document collections consist of candidate codes.}
\label{tab:csndata}
\end{center}
\end{table}

\subsection{Additional Evaluation Results of SANTA}\label{app:CSN}
In this experiment, we follow~\citet{li2022coderetriever}, keep the same evaluation settings and evaluate the retrieval effectiveness of SANTA on CodeSearch dataset. The dataset consists of code retrieval tasks on six programming languages, including Ruby, Javascript, Go, Python, Java, and PHP. We show the data statistics of CodeSearch in Table~\ref{tab:csndata}. Since CodeT5 and CodeRetriever don't release their data processing code for pretraining. We can only refer to the tutorial\footnote{\url{https://github.com/github/CodeSearchNet/blob/master/notebooks/ExploreData.ipynb}} to process data. When we evaluate SANTA on CodeSearch, the instances in testing and development sets are filtered out from CodeSearchNet dataset for pretraining. Some codes that can not be parsed are also filtered out, because the data processing details are not available\footnote{\url{https://github.com/salesforce/CodeT5/issues/64}}.
%Different from the setting in~\cite{DBLP:journals/corr/abs-1909-09436}, it has filtered low-quality queries by handcrafted rules and expanded 1000 candidates to the whole code corpus.

During continuous pretraining, we set the batch size, learning rate and epoch as 128, 5e-5 and 10, respectively. During finetuning, we set the learning rate as 2e-5 and 1e-5 for CodeSearch and Adv, and set batch size and epoch as 128 and 12. We use inbatch negatives with one hard negative for finetuning and the hard negative is randomly sampled from the top 100 retrieved negative codes by pretrained SANTA. The warm-up ratio is 0.1.

The performance of SANTA on CodeSearch and Adv is shown in Table~\ref{tab:CSN}. Under the zero-shot setting, SANTA still outperforms CodeRetriever~\cite{li2022coderetriever} with about 2\% improvements, which shows that the advances of SANTA can be generalized to different structured data retrieval tasks. Moreover, SANTA also shows strong zero-shot ability by achieving comparable performance with the finetuned CodeBERT, GraphCodeBERT and CodeT5 models. After finetuning, SANTA achieves more than 3.7\% improvements over CodeT5 on CodeSearch. All these encouraged experiment results further demonstrate that our structure-aware pretraining method indeed helps language models to capture the structure semantics behind the text data. The retrieval performance on Adv dataset illustrates that the retrieval effectiveness of SANTA can be further improved by increasing the batch size from 16 to 128.

\end{appendix}

\end{document}